\documentclass[12pt]{article}
\def\Journal#1#2#3#4{{#1} {\bf #2}, #3 (#4)}

\def\NPB{{\em Nucl. Phys.} B}
\def\PLB{{\em Phys. Lett.}  B}
\def\PRL{\em Phys. Rev. Lett.}
\def\PRD{{\em Phys. Rev.} D}

\def\be{\begin{equation}}
\def\ee{\end{equation}}
\def\bea{\begin{eqnarray}}
\def\eea{\end{eqnarray}}

\begin{document}
\title{Some Considerations on Chiral Gauge Theories}
\author{Massimo Testa \\
Dip. di Fisica, Universit\'{a} di Roma
``La Sapienza``\\ and\\ INFN Sezione di Roma I\\
Piazzale Aldo Moro 2\\I-00185 Roma\\
 Italy\\ massimo.testa@roma1.infn.it}
\maketitle
\begin{abstract}
Some general considerations
on the problem of non-perturbative definition of
Chiral Gauge Theories are presented.
\end{abstract}
%
\section{Introduction}

While, classically, Vector and Chiral Gauge Theories look quite similar, 
their quantization proceeds through quite different lines, due to the 
phenomenon of chiral anomalies, which is reflected in the lack of a chiral
invariant regularization.

A non-perturbative formulation of Chiral Gauge Theories could clarify 
fundamental issues, as the possibility of dynamical Higgs mechanism, 
baryon non conservation and the question of naturalness.

How should we quantize chiral gauge theories?

Several approaches have been explored:
\begin{enumerate}
\item Non gauge invariant
quantization{\cite{borrelli1},\cite{borrelli2}},\cite{romchir}
(Rome approach\footnote{Within this class falls
also the formulation of the Zaragoza 
group\cite{alonso}.}) based on the Bogolubov
method\footnote{Recent work in the framework of the Rome approach,
including numerical simulations, has been
carried on by the authors of ref.\cite{bgs}.}

\item Gauge invariant quantization\cite{smit},\cite{aoki}

\item Mirror Fermions\cite{montvay}

\item Fine-Grained Fermions\cite{sch},\cite{hoo},\cite{hs}

\item Other dimensions or infinite number of regulators\cite{kaplan},\cite{Slavnov}

\item Overlap\cite{nn},\cite{n1},\cite{rds}

\item Ginsparg-Wilson Fermions\cite{G-W},\cite{luscher},\cite{hasenfratz}
\end{enumerate}

The Overlap and the Ginsparg-Wilson approach make in fact use of the 
same fermion discretization, while differ in the treatment of the 
residual gauge invariance breaking, present at the lattice level.

In the following I will present some general considerations on 
the problem of quantization of Chiral Gauge Theories and on
some of its solutions presented so far.

\section{Target Theory}

The first step, common to every approach, is to define the 
target theory we want to reproduce in the continuum. The target 
theory is usually identified through its symmetries and is defined 
by a formal, continuum, Lagrangian density:
\begin{equation}
L_g=\bar \psi _L\not D\psi _L+{1 \over 4}W_{\mu \nu }^aW_{\mu \nu 
}^a\label{uno}
\end{equation}
where:
\begin{equation}
D_\mu \equiv \partial _\mu +ig_{0}W_\mu ^a\,T^a
	\label{due}
\end{equation}
In eq.(\ref{due}), the $T^{a}$'s are the
appropriate generators of the  gauge group $G$ and
$g_{0}$ denotes the bare coupling.
Higgs degrees of freedom could easily be added to the action in eq.(\ref{uno}),
but for simplicity we will not do so.

The general problem of the quantization of chiral theories is well known:
regularization is, in general, incompatible with exact chiral 
invariance. This incompatibility is unavoidable and is at the origin of
anomalies.

For example, naive discretization of Dirac action suffers from the 
so-called Doubling Problem: the spectrum of naively discretized
fermions is vector-like rather than chiral.

A possible solution to this problem consists in the introduction of the 
so-called Wilson term\cite{wilson}:
\begin{equation}
L_W\approx r\bar \psi _La\partial ^2\psi _R+h.c.
	\label{tre}
\end{equation}
or, in the Wilson Majorana form\cite{redundant},
\begin{equation}
L_W=a\left( {\chi ^\alpha \partial _\mu \partial _\mu \chi _\alpha 
+\bar \chi ^{\dot \alpha }\partial _\mu \partial _\mu \bar \chi 
_{\dot \alpha }} \right)	
	\label{quattro}
\end{equation}
where $\chi ^\alpha(x)$ and $\bar \chi ^{\dot \alpha }(x)$ are
bispinor Grassmann fields.
$L_W$ is a chiral violating\cite{nielseninomiya}, ''irrelevant'' term: its presence can, and 
must, be compensated by finite or
power divergent counterterms. No new logarithms appear as will be 
discussed later on.
The Overlap\cite{nn},\cite{n1} and the Ginsparg-Wilson\cite{luscher}
formulation are based on a different discretization which captures more
geometrical 
meaning with respect to the Dirac-Wilson one, but is confronted with a similar problem: 
the unmodified action $S$ of an anomaly free chiral gauge  theory has the form:
\begin{equation}
S=S_{GI}+a\int {d^4xO_5(x)}	
	\label{cinque}
\end{equation}
where $O_5(x)$ is a non gauge invariant operator of dimension $5$.
As it will be discussed in the following sections, the presence of an
uncompensated $aO_5$ term is, potentially, very dangerous.

\section{Uncompensated Symmetry Breaking}

Let us consider, e.g., a $\lambda \phi ^4$ theory symmetric under
$\phi \leftrightarrow -\phi $.
As an example of what happens if the symmetry breaking 
induced by the regulator is not compensated by the introduction of 
appropriate counterterms, we may study the consequences of adding to 
the Lagrangian density
an irrelevant term $a O_5\propto a \phi ^5$, which violates this symmetry.

In order to be really a correction of order $a$, $O_5$
should be multiplicatively renormalizable in the continuum limit 
($a\to 0$), in order to avoid an immediate back-reaction, through 
mixing, giving rise to terms $O_3\approx \phi ^3$ and
$O_1\approx \phi $. It will consist, therefore, of an appropriate 
linear combination of $\phi ^5$, $\phi ^3$ and $\phi $.

We will study the resulting theory by expanding Green's functions
in powers of $a\int {dx\,O_5}$, while keeping the dependence on the coupling
constant completely non-perturbative.
Let us consider, in particular, the three-point Green's function 
which is expected 
to vanish in the symmetric $\phi \leftrightarrow -\phi $ theory:
\begin{equation}
\left\langle {\phi (x_1)\phi (x_2)\phi (x_3)} \right\rangle 
=\sum\limits_{n=0}^\infty  {{1 \over {n!}}\left\langle {\phi 
(x_1)\phi (x_2)\phi (x_3)(a\int {dyO_5(y))^n}} \right\rangle }
	\label{sei}
\end{equation}
We will now discuss what happens at various orders in the insertion of 
$a\int {dx\,O_5}$.

$\bullet$ First order correction
\begin{equation}
\left\langle {\phi (x_1)\phi (x_2)\phi (x_3)} \right\rangle 
_{(1)}=a\int {dy\left\langle {\phi (x_1)\phi (x_2)\phi (x_3)O_5(y)} 
\right\rangle }	
	\label{sette}
\end{equation}
This is O.K. (it vanishes as
${a\to 0}$)
since $O_5$ is multiplicatively renormalizable and, therefore, its single
insertion is only logarithmically divergent.

$\bullet$ Higher order corrections

The next interesting contribution, in this particular example, is 
the third order insertion:
\begin{equation}
	\left\langle {\phi (x_1)\phi (x_2)\phi (x_3)} 
\right\rangle _{(3)}=a^3 \int {dy_1 dy_2 dy_3 \left\langle {\phi (x_1)\phi (x_2)
\phi(x_3) O_5(y_1) O_5(y_2) O_5(y_3)} \right\rangle}
	\label{otto}
\end{equation}
The r.h.s of eq.(\ref{otto}), although formally of order $a^3$, is
in fact divergent and violates the symmetry
$\phi \leftrightarrow -\phi $.

This follows from the fact, in eq.(\ref{otto}), the contribution from the
integrals over $y_1$, $y_2$, $y_3$,
when the $y$'s are close together, is enhanced by short 
distance operator singularities, and can be estimated through
power-counting\footnote{Eq.(\ref{nove}) is meant as an illustration 
only. 
In the r.h.s. there will also be a contribution from $O_{5}$, which 
will not survive in eq.(\ref{tredici}) and a contribution from 
$O_{1}$, even more singular, which obviously leads to the same 
qualitative conclusions.}:
\begin{equation}
\int {dy_1dy_2dy_3O_5(y_1)O_5(y_2)O_5(y_3)}
\mathrel{\mathop{\kern0pt\approx}\limits_{a\to 0}} {c \over 
{a^4}}\int {dyO_3(y)}	
	\label{nove}
\end{equation}
with an appropriately chosen $c$.

Using the Renormalization Group we will now sketch the proof that $c$ can only depend
on $g_{0}$\cite{broken}.
In fact, if we denote by $ Z_5$ the logarithmically 
divergent renormalization constant which makes the single insertion
of $O_{5}$ finite, we have that
the triple composite operator insertion:
\begin{equation}
I(y_1,y_2,y_3) \equiv Z_5^3O_5(y_1)O_5(y_2)O_5(y_3)	
	\label{dieci}
\end{equation}
is finite (as $a\to 0$), but not integrable, i.e. it is not a 
distribution and $\int {dy_1dy_2dy_3I(y_1,y_2,y_3)}$
does not exist in the continuum.
$I(y_1,y_2,y_3)$ may, however, be transformed into a distribution through 
an appropriate subtraction such that:
\begin{equation}
T\equiv Z_5^3(\int {dy_1dy_2dy_3O_5(y_1)O_5(y_2)O_5(y_3)}-{c \over 
{a^4}}\int {dyO_3(y)})	
	\label{undici}
\end{equation}
stays finite when $a\to 0$, when inserted in a particular Green 
function.

$c$ is dimensionless and may, a priori, depend on $g_0$ and $a \mu$:
\begin{equation}
c=c(g_0,a\mu)
\end{equation}

We then have:
\begin{equation}
\int {dy_1dy_2dy_3O_5(y_1)O_5(y_2)O_5(y_3)}={1 \over {Z_5^3}}T+{c 
\over {a^4}}\int {dyO_3(y)})	
	\label{dodici}
\end{equation}
so that:
\begin{eqnarray}
a^3\int {dy_1dy_2dy_3O_5(y_1)O_5(y_2)O_5(y_3)}=
{{a^3} \over {Z_5^3}}T+{c \over a}\int {dyO_3(y)}) \approx \nonumber \\
\mathrel{\mathop{\kern0pt\approx}\limits_{a\to 0}} {c \over a}\int {dyO_3(y)})	
	\label{tredici}
\end{eqnarray}

Let us now consider the Callan-Symanzik differential operator\cite{cal}:
\begin{equation}
\left. {\mu {d \over {d\mu }}} 
\right|_{g_0,a}\equiv \mu {\partial  \over {\partial \mu }}+\beta 
(g){\partial  \over {\partial g}}	
	\label{quattordici}
\end{equation}
 
As evident from eq.(\ref{quattordici}), $\left. {\mu {d \over {d\mu }}} 
\right|_{g_0,a}$ transforms u.v. finite quantities into finite
quantities. Applying it to $T$, eq.(\ref{undici}), we 
get\footnote{Eq.(\ref{quindici}) must be considered as a symbolical
equation. The complete argument would require the 
insertion of $T$ into a Green function. For a thorough treatment see 
ref.\cite{broken}.}:
\begin{eqnarray}
\left. {\mu {{dT} \over {d\mu }}} \right|_{g_0,a}
=3\mu \left. {{{d\log Z_5} \over {d\mu }}} \right|_{g_0,a}T-Z_5^3{{\int 
{dy_1O_3(y_1)}} \over {a^4}}\left. {\mu {{dc} \over {d\mu }}} 
\right|_{g_0,a}=\nonumber \\ 
=3\gamma _{O_3}T-Z_5^3{{\int {dy_1O_3(y_1)}} \over {a^4}}\left. 
{\mu {{dc} \over {d\mu }}} \right|_{g_0,a} \label{quindici}
\end{eqnarray}
In eq.(\ref{quindici}), when $\left. {\mu {d \over {d\mu }}} \right|_{g_0,a}$
acts on bare, unrenormalized quantities like
$O_5(y_1)O_5(y_2)O_5(y_3)$ and $\int {dyO_3(y)}$, gives $0$
because they only depend from $a$ and $g_{0}$ . 

Eq.(\ref{quindici}) shows that
$\left. {\mu {{dT} \over {d\mu }}} \right|_{g_0,a}$ is u.v. finite only if:
\begin{equation}
\left. {\mu {{dc} \over {d\mu }}} \right|_{g_0,a}=0	
	\label{sedici}
\end{equation}
i.e., if $c$ is a function of $g_0$ only:
\begin{equation}
c=c(g_{0})
	\label{diciassette}
\end{equation}

The integration region where the $y$'s are close together, which is the 
only one surviving the continuum limit in eq.(\ref{otto}), gives 
therefore rise to
a linearly divergent contribution:
\begin{equation}
\left\langle {\phi (x_1)\phi (x_2)\phi (x_3)} \right\rangle _{(3)}
\approx {c(g_{0}) \over a}\int {dy}\left\langle {\phi (x_1)\phi 
(x_2)\phi (x_3)O_3(y)} \right\rangle	
	\label{diciotto}
\end{equation}
with a coefficient depending on the bare parameters only.

Depending on the particular regularization employed, the appearance 
of these unwanted contributions can be shifted to higher orders in 
perturbation theory, but can hardly be eliminated, unless some exact 
selection rule is operating at the level of the regularized theory.

In order to treat the similar problem arising in Chiral Gauge 
Theories, the following strategies have been proposed:

\begin{itemize}
\item Add to the action compensating counterterms\footnote{In
order to show in a completely non-perturbative way that the 
adjustement of the coefficients of \underline {local} counterterms in 
the action allows the elimination of such unwanted contributions, one 
should show that the symmetry breaking insertions exponentiate. This 
can be done following a line of reasoning similar to the one
illustrated in sect.\ref{gauge}.};

\item Gauge average
\end{itemize}

In the first class are included, although with quite a different 
status, the Rome approach\cite{borrelli1} and the L\"{u}scher\cite{luscher}
approach, based on the Ginsparg-Wilson\cite{G-W} discretization. The use 
of Ginsparg-Wilson fermions allows, as shown by 
L\"{u}scher\cite{luscher}, the very interesting possibility of
compensating the lack of gauge invariance through the addition of appropriate
counterterms, thus obtaining a gauge invariant theory at \underline {finite} 
lattice spacing. Through this procedure it is possible to avoid
the necessity of a gauge fixing term, which is
peculiar of the Rome approach.

Neuberger proposal\cite{n1}, instead, is based on the conjecture 
that the breaking of gauge invariance takes care of itself, through 
averaging, according to a mechanism first suggested in ref.\cite{foerster}. 

\section{Rome Approach}

In this section I will recall the basic strategy followed within the 
Rome approach\cite{borrelli1}.

One starts defining the vector fields through the link variables as e.g.:
\begin{equation}
ag_0W_\mu \equiv {{U_\mu -U_\mu ^+} \over {2i}}
	\label{diciannove}
\end{equation}
and then adds to the action the discretization of the continuum gauge-fixing
term:
\begin{equation}
L_{gf}={1 \over {2\alpha _0}}(\partial _\mu W_\mu ^a)(\partial _\mu 
W_\mu ^a)+\bar c\partial _\mu D_\mu c 	
\label{venti}
\end{equation}
The gauge-fixing Lagrangian in eq.(\ref{venti}) can be linearized through
a set of Lagrange multipliers $\lambda^{a}(x)$, as:
\begin{equation}
L_{gf}={{\alpha _0} \over 2}\lambda ^a\lambda ^a+i\lambda 
^a(\partial _\mu W_\mu ^a)+\bar c\partial _\mu D_\mu c
	\label{ventuno}
\end{equation}
Beside the term written in eq.(\ref{ventuno}), one also adds all possible
symmetry breaking (and non-Lorentz invariant in the 
case of the Lattice Discretization) counterterms with dimension $D\le 
4$. As examples we mention:
\begin{eqnarray}
-{{\delta \mu _W^2} \over 2}W_\mu ^a(x)W_\mu ^a(x) \label{mass} \\
\delta Z_{1}\left( {\partial _\mu W_\mu ^a} \right)^2 \label{dless1} \\ 
\delta Z_{2} f^{abc}\partial _\mu W_\nu ^aW_\mu ^bW_\nu ^c  
\label{dless2} \\
\delta Z_{3} \sum\limits_\mu  {\partial _\mu W_\mu ^a\partial _\mu W_\mu 
^a}	 \label{dless3} \\
\delta g_{gh}f_{abc}\bar c^a\partial _\mu \left( {W_\mu ^bc^c} 
\right) \label{ghost}
\end{eqnarray}
The presence of the counterterm with $\delta g_{gh}$, eq.(\ref{ghost}),
signals an irreversible breaking of gauge geometry, within this 
approach, as discussed later.

The target gauge fixed theory is invariant under the BRST 
symmetry\cite{brst}:
\begin{eqnarray}
\delta \chi^\alpha =\varepsilon \delta _{BRST}\chi^\alpha 
=i\varepsilon g_0c^aT^a\chi ^\alpha	\nonumber \\
\delta W_\mu ^a\equiv \varepsilon \delta _{BRST}W_\mu ^a=\varepsilon 
(D_\mu c)^a \nonumber \\
\delta \lambda^a=0 \label{ventidue} \\
\delta c^a\equiv \varepsilon \delta _{BRST}c^a=-{1 \over 
2}\varepsilon g_0f_{abc}c^bc^c \nonumber \\
\delta \bar c^a\equiv \varepsilon \delta _{BRST}\bar c^a=\varepsilon 
\;i\lambda^a \nonumber
\end{eqnarray}
As well known these transformations are nilpotent:
\begin{equation}
\delta_{BRST}^2=0	
	\label{ventitre}
\end{equation}
so that the target gauge fixing action, eq.(\ref{ventuno}), which can be 
rewritten as
\begin{equation}
L_{gf}={{\alpha _0} \over 2}\lambda ^a\lambda ^a+\delta _{BRST}(\bar 
c^a\partial _\mu W_\mu ^a)	
	\label{ventiquattro}
\end{equation}
is BRST-invariant:
\begin{equation}
\delta _{BRST}L_{gf}=0	
	\label{venticinque}
\end{equation}
BRST invariance implies, in the target theory,
an infinite set of identities 
on the Green Functions:
\begin{equation}
\left\langle {\Phi _1(x_1)\ldots \ldots \Phi _n(x_n)} \right\rangle 
\equiv \int {d\mu \;e^{-S_{cl}}\Phi _1(x_1)\ldots \ldots \Phi 
_n(x_n)}	
	\label{ventisei}
\end{equation}
as:
\begin{equation}
\left\langle {\delta _{BRST}\left( {\Phi _1(x_1)\ldots \ldots \Phi 
_n(x_n)} \right)} \right\rangle =0	
	\label{ventisette}
\end{equation}
The strategy, then, is to fix the values of the counterterms as follows.
We compute, first of all, non-perturbatively:
\begin{eqnarray}
\left\langle {\Phi _1(x_1)\ldots \ldots \Phi _n(x_n)} 
\right\rangle = \nonumber \\
=\int {DU_\mu D\bar \chi D\chi D\bar cDc} \label{ventotto} \\
e^{-S_0+S_W+{1 \over {2\alpha _0}}\int {d^4x\left( {\partial _\mu 
W_\mu ^a} \right)^2+S_{ghost}+S_{c.t.}}}\Phi _1(x_1)\ldots \ldots 
\Phi _n(x_n) \nonumber
\end{eqnarray}
and tune the values of the counterterms by imposing the 
BRST-"Identities'', eq.(\ref{ventisette}).
This is at best possible up to order $a$, and impossible if there are 
unmatched anomalies.

This procedure defines a ÒbareÓ chiral theory with parameters $g_0$
 and $\alpha _0$ defined by the BRST transformations.
We can then carry out the usual non-perturbative renormalization, by 
fixing the independent bare parameters $g_0$ and $\alpha _0$ to 
reproduce given finiteness conditions on physical quantities and/or 
Green functions.

I want to stress that this procedure is completely non-perturbative, 
but can also be checked in Perturbation 
Theory\cite{borrelli1},\cite{trav}.
Due to the rather large number of possible counterterms, the
perturbative evaluation of some of them might be helpful in practice.
In fact the theory so defined should be asymptotically free and we may 
distinguish two different kinds of counterterms:

$\bullet$	Dimensionless counterterms:
\begin{equation}
\delta Z=f_Z(g_0,\alpha _0)	
	\label{ventinove}
\end{equation}
like those appearing in 
eqs.(\ref{dless1})-(\ref{ghost}).

$\bullet$	Dimensionful counterterms:
\begin{equation}
\delta M={{f_M(g_0,\alpha _0)} \over a}	
	\label{trenta}
\end{equation}
like the one in eq.(\ref{mass}).
While dimensionless counterterms can be safely estimated in 
perturbation theory, since $g_{0} \rightarrow 0$, the dimensionful 
counterterms are essentially non-perturbative.
In fact exponentially small contributions to $f_M$ can be rescued by ${1 
\over a}$:
\begin{equation}
{{f(g_0,\alpha _0)} \over a}\approx {{e^{-\,{1 \over {g_0^2}}}} 
\over a}\approx \Lambda 	
	\label{trentuno}
\end{equation}
where $\Lambda$ is a non-perturbative parameter of the same nature as 
$\Lambda_{QCD}$.

To conclude this section, I would like to add that ghosts and gauge-fixing
are unavoidable features of the Rome approach. In 
fact in a two loop computation in dimensional 
regularization\cite{rossi} the ghost counterterm, eq.(\ref{ghost}),
has been shown to be present, so that, at least if we trust Perturbation
Theory, we cannot invert the 
Faddeev-Popov procedure and remove the gauge fixing.

\section{Gauge-Average without Gauge-Fixing} \label {gauge}

Another possible way to cope with $O(a)$ violations of chiral 
gauge invariance due to the regulator, consists in the use
of the gauge non-invariant theory\cite{foerster},\cite{smit},\cite{aoki},\cite{n1}:
\begin{equation}
S=S_{GI}+a\int {d^4yO_{5}(y)}	
	\label{trentadue}
\end{equation}
without any gauge-fixing\footnote{We are using continuum notations for
the lattice regulated theory.}. According to the Nielsen-Ninomiya 
theorem\cite{nielseninomiya}, the theory described by $S_{GI}$ 
alone would be vector-like because of the presence of doublers.

The basic idea behind the strategy of gauge-average is the 
possibility that the gauge invariant integration
measure in the functional integral, will automatically take care of the
''small'' violations induced by the regulator, leading, in the end, to 
the correct theory. This mechanism is essentially non-perturbative 
and, therefore, it stands like a conjecture, difficult to prove and difficult to
disprove.
In the following I will present some heuristic remarks about this 
conjecture.

For an observable, gauge invariant quantity, $O_{GI}$, the procedure of 
using eq.(\ref{trentadue}) without any gauge-fixing, is in fact equivalent
to an average over gauge transformations $\Omega $:
\begin{eqnarray}
\left\langle {O_{GI}} \right\rangle \equiv \int {DUD\bar \chi D\chi }e^{S_{GI}+
a\int {dyO_{5}(y)}}O_{GI}= \nonumber \\
=\int {DUD\bar \chi D\chi D\Omega}e^{S_{GI}+a\int 
{dyO_{5}(y)^{\Omega}}}O_{GI} \label{trentatre}
\end{eqnarray}
Since the theory is lattice regulated, the $\Omega$-integration
in eqs.(\ref{trentatre}) is compact and obeys the
rules of group theory, as, e.g.:
\begin{equation}
\int {D\Omega }\,\Omega _{ij}(x)\Omega _{kl}^+(y)=\delta _{xy}\delta 
_{il}\delta _{jk}	
	\label{trentaquattro}
\end{equation}
We have now to study the general form of the insertion $\int{D\Omega e^{a\int 
{dyO_{5}(y)^{\Omega}}}}$ in eq.(\ref{trentatre}). To do this we 
expand the integrand in powers of $a\int{dy O_{5}(y)^{\Omega}}$:
\begin{eqnarray}
\left\langle {O_{GI}} \right\rangle= \sum\limits_{n=0}^\infty  {\left\langle {O_{GI}} 
\right\rangle _{(n)}=}\sum\limits_{n=0}^\infty  {{{a^n} \over {n!}}\int {D\Omega 
}\left\langle {O_{GI}
(\int {d^4yO_{5}(y)^\Omega })^n} \right\rangle} \label{trentatreb}
\end{eqnarray}
The $\Omega$ integration in eq.(\ref{trentatreb}) projects the gauge invariant
contribution out of $O_{5}(y)$ and their products. In virtue of eq.(\ref{trentaquattro})
a string of $O_{5}(y)$'s will, after gauge average, cluster into a sum 
of local gauge 
invariant operators multiplied by space-time $\delta$-functions.

As an example, let us first consider a simplified situation in 
which $\int{D\Omega O_{5}(x)^{\Omega}}=0$ and
$\int{D\Omega O_{5}(x)^{\Omega} O_{5}(y)^{\Omega}}=F(x) \delta(x-y)$ with a 
\underline{local}
gauge invariant $F(x)$. In this case it is easy to resum the series 
in eq.(\ref{trentatreb}):
\begin{eqnarray}
\int{D\Omega e^{a\int{dyO_{5}(y)^{\Omega}}}}=
{1\over 2!}a^{2}\int{dx_{1} dx_{2}
\int{D\Omega O_{5}(x_{1})^{\Omega} O_{5}(x_{2})^{\Omega}}}+ \nonumber \\
+{1\over 4!}a^{4}\int{dx_{1}\dots dx_{4}
\int{D\Omega O_{5}(x_{1})^{\Omega}\dots O_{5}(x_{4})^{\Omega}}}+\dots 
=\nonumber\\
=\int{dx_{1} a^{2} {F(x_{1})\over 2}}+{1\over 2!} \int{dx_{1} dx_{2}
a^{2}{F(x_{1})\over 2}a^{2} {F(x_{2})\over 2}}+\dots= \nonumber \\
=e^{\int{dy a^{2}{F(y)\over 2}}}
\label{series}
\end{eqnarray}
In the general case the structure displayed in eq.(\ref{series}) is 
maintained, with $a^{2}{F(y)\over 2}$ replaced by an infinite series 
of irrelevant, local, gauge invariant operators obtained by projecting
the gauge invariant contribution out of the product of an increasing number of $O_{5}$'s. 
Although local, these operators have an increasing spread in $y$ and
the question is if they can really be treated as irrelevant insertions. 
If they could, then, as far as the expectation value of gauge 
invariant observables is 
concerned, the theory would be in the same universality 
class as the one with the exactly gauge invariant action $S_{GI}$ in eq.(\ref{trentadue}). 
Therefore the doublers would be back and we would end up with
a Vector-like Spectrum. The answer to this question, as remarked before, lies 
outside the scope of perturbation theory. In particular there could 
exist different fixed points, with different anomalous dimensions,
corresponding to different continuum 
theories. Although conceivable, this possibility looks rather unlikely 
when, as in the present case, the theory to be defined is asymptotically
free and therefore controlled by the fixed point at zero coupling.

\section{Conclusions}

The quantization of Chiral Gauge Theories is still a fascinating and 
evolving subject. We are witnessing very interesting progress, with 
possible applications also in the framework of consolidated vector 
theories as, for instance, QCD.

It must be remarked, however, that even these new developments do not 
put Chiral and Vector theories on the same ground and seem to maintain
some basic, deep difference, in 
particular in what concerns the subjects of fine tuning and naturalness.

\section*{Acknowledgements}
I thank the organizers of the Chiral `99 Workshop and in particular
Professor Ting-Wai Chiu for the warm 
hospitality and for providing a very fruitful and stimulating atmosphere.

\end{document}